\documentclass{article}
\usepackage{epsf}

\def\res{\;\raisebox{-2.5mm}{\epsfysize=2cm\epsfbox{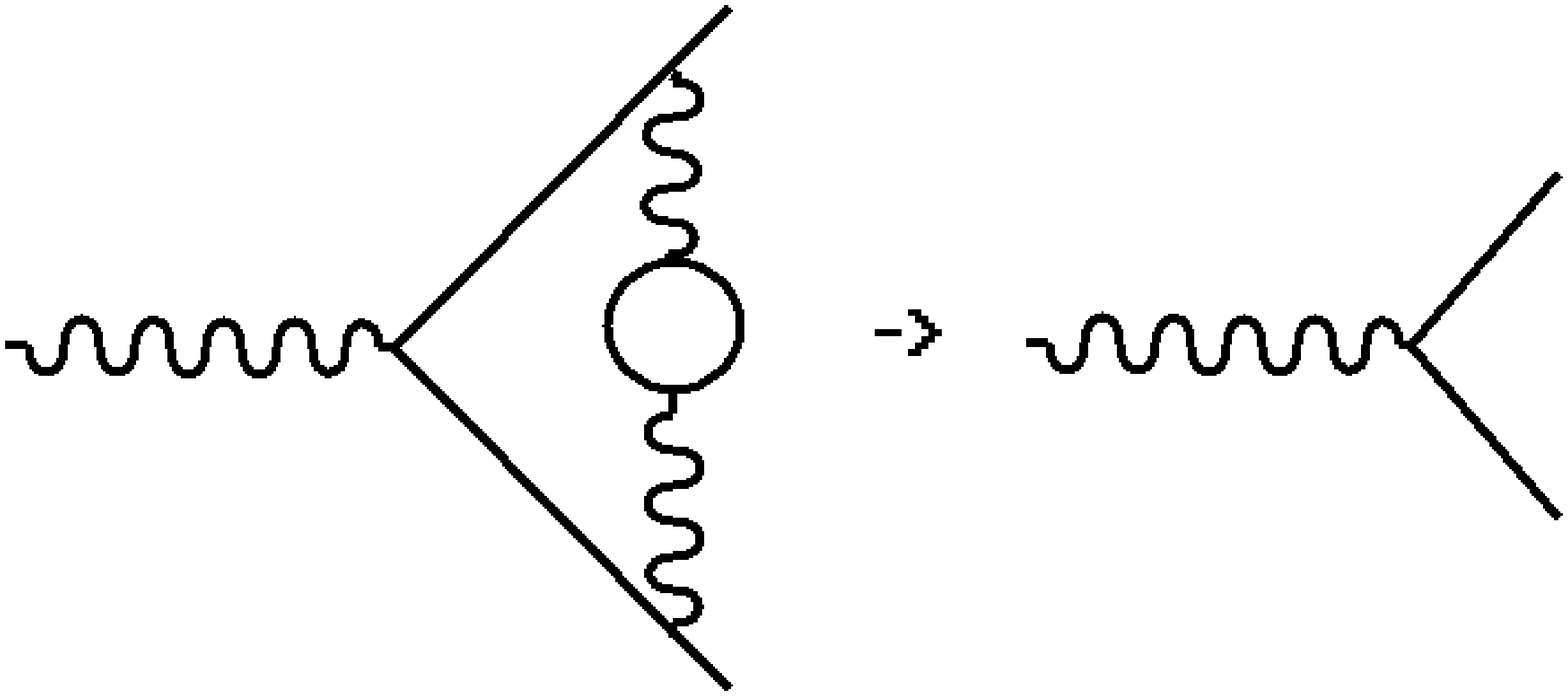}}\;}
\def\pica{\;\raisebox{-2.5mm}{\epsfysize=4cm\epsfbox{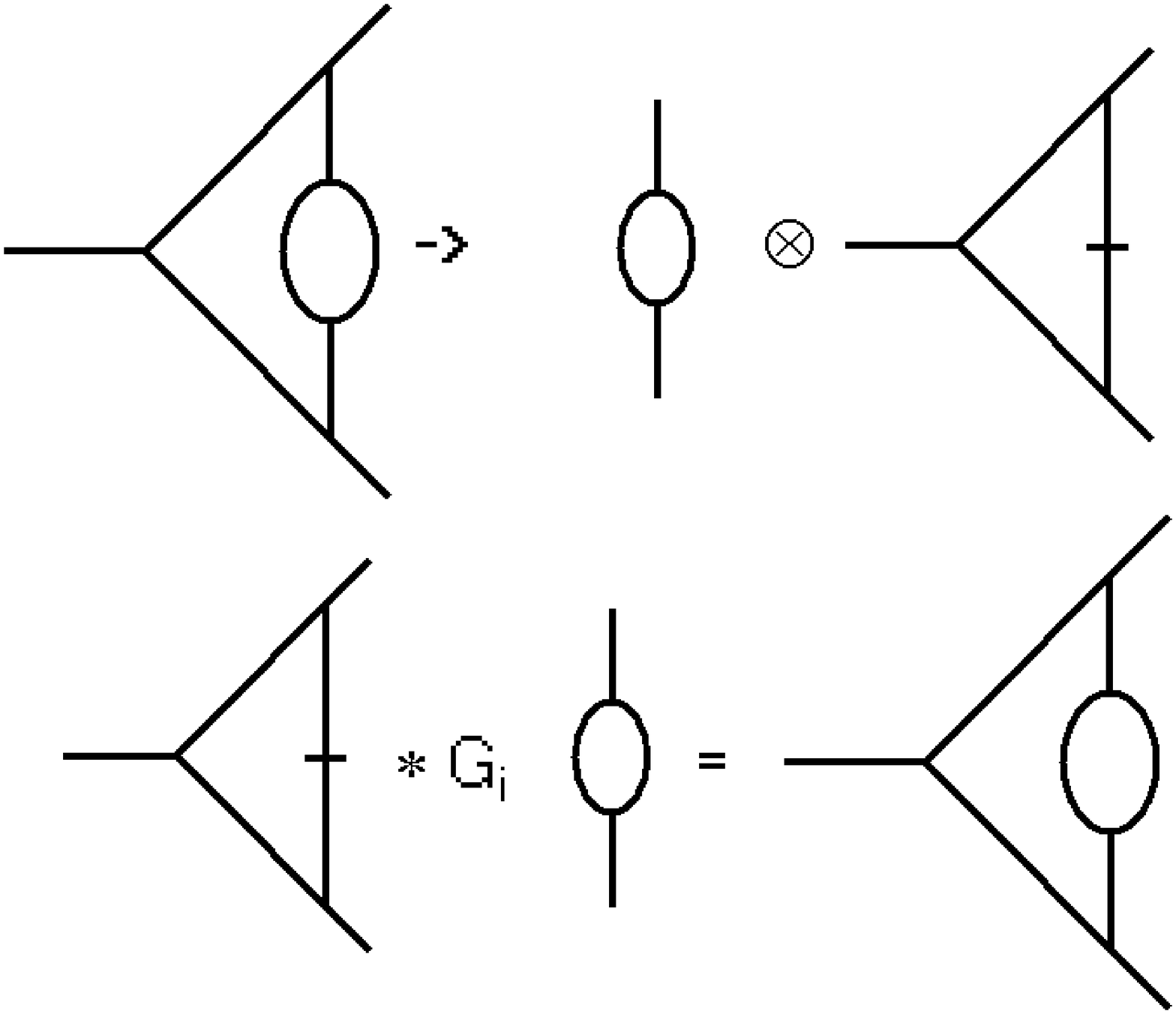}}\;}
\def\picb{\;\raisebox{-2.5mm}{\epsfysize=4cm\epsfbox{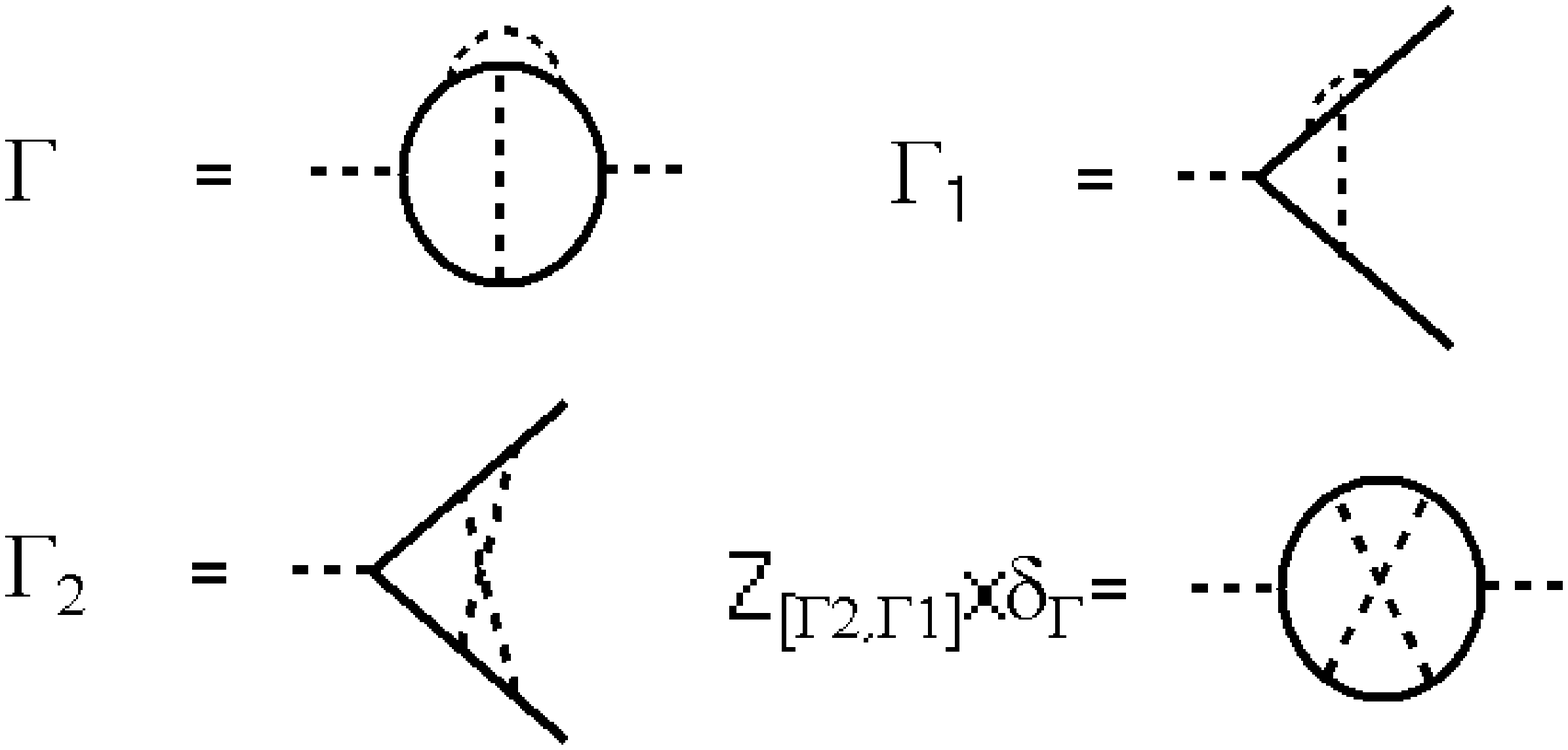}}\;}
\def\picf{\;\raisebox{-2.5mm}{\epsfysize=4cm\epsfbox{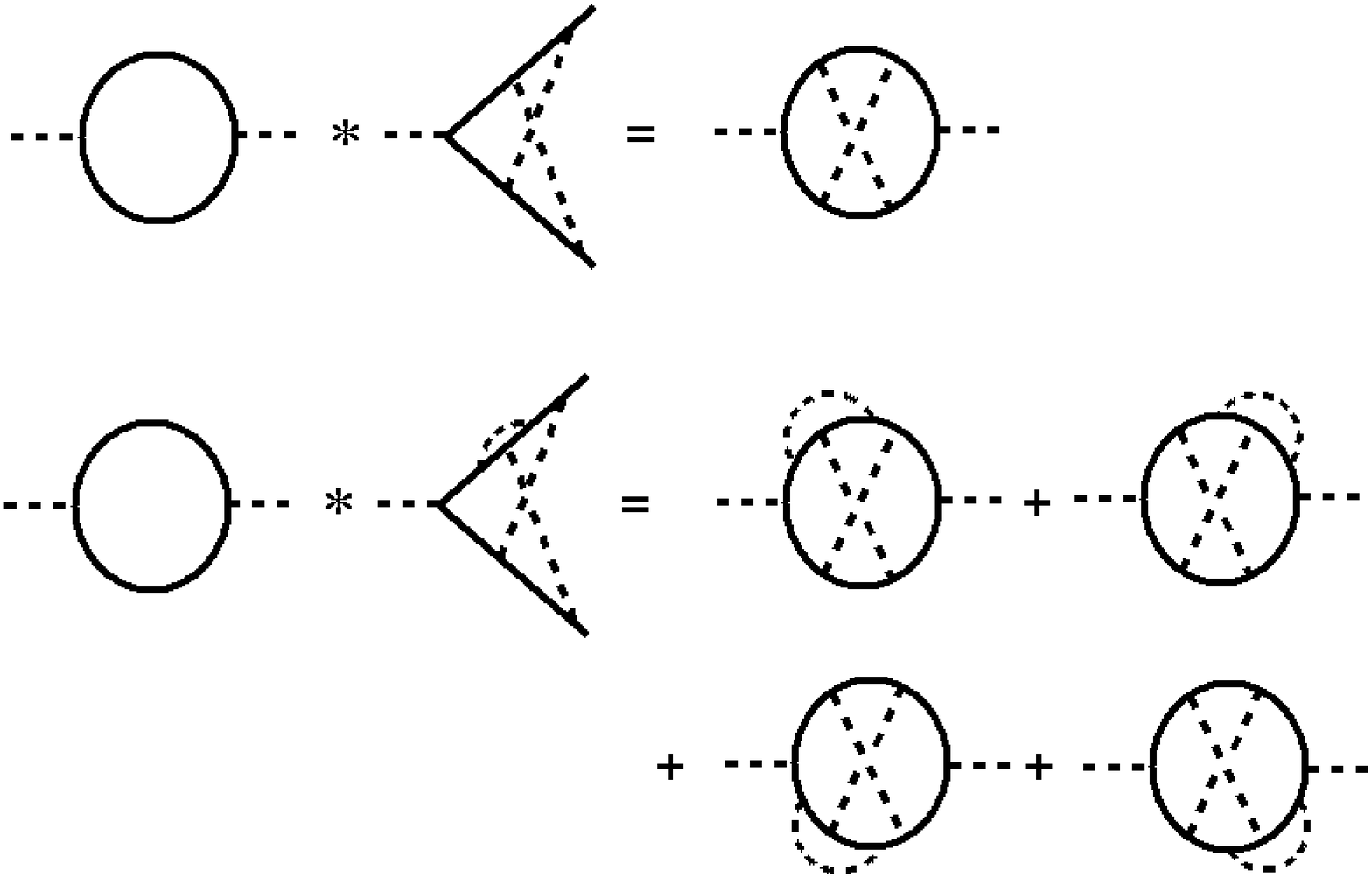}}\;}
\def\picgt{\;\raisebox{-2.5mm}{\epsfysize=4cm\epsfbox{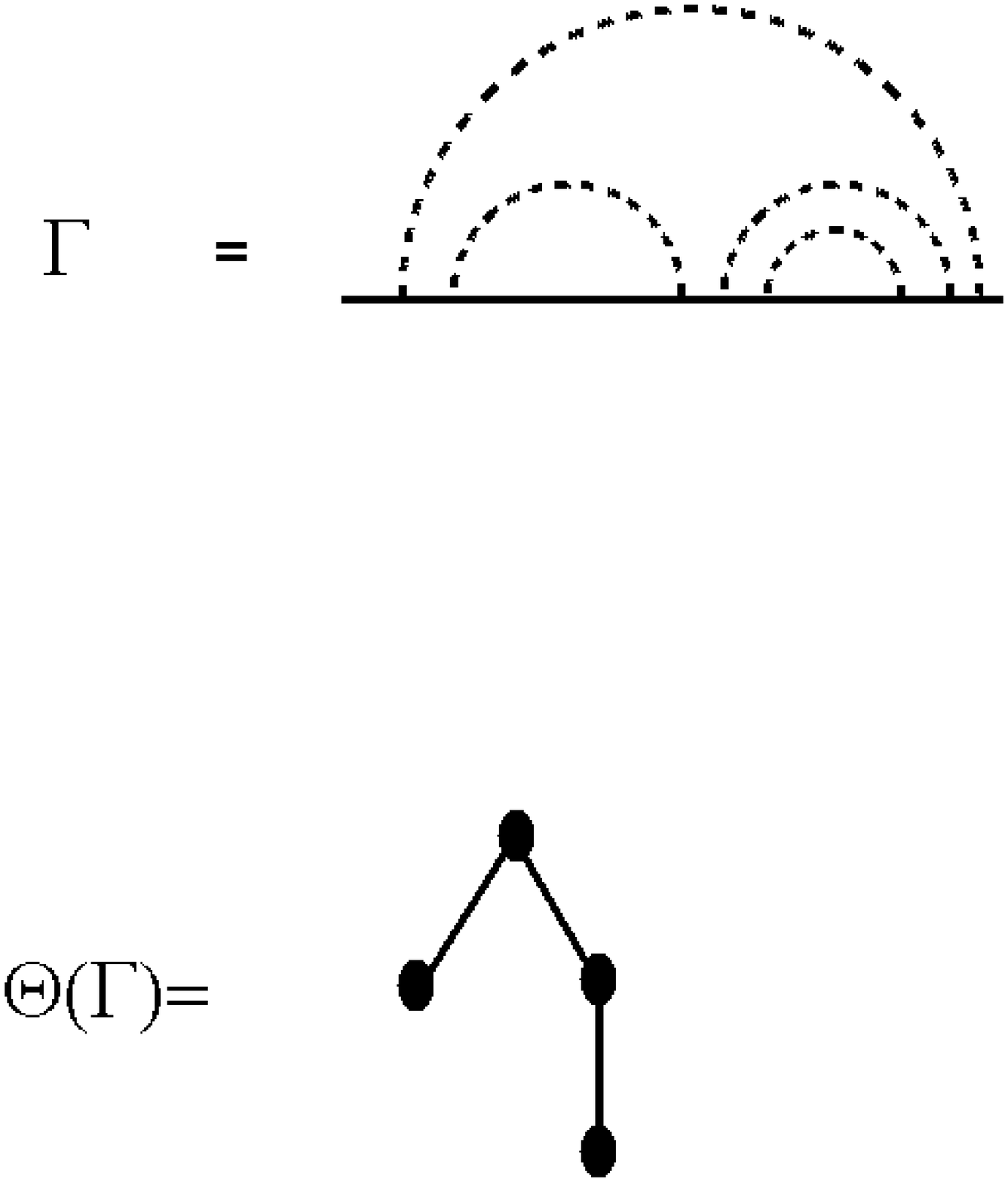}}\;}
\def\olp{\;\raisebox{-1.2mm}{\epsfysize=3mm\epsfbox{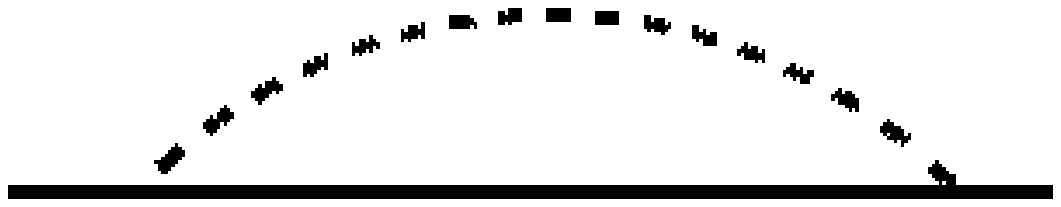}}\;}
\def\srt{\;\raisebox{-1.3mm}{\epsfysize=4mm\epsfbox{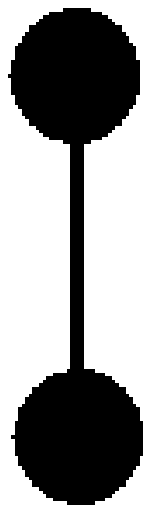}}\;}

\def\build#1_#2^#3{\mathrel{
\mathop{\kern 0pt#1}\limits_{#2}^{#3}}}

\font\tenbb=msbm10 \font\sevenbb=msbm7 \font\fivebb=msbm5
\newfam\bbfam
\textfont\bbfam=\tenbb \scriptfont\bbfam=\sevenbb
\scriptscriptfont\bbfam=\fivebb

\begin{document}
\title{\bf Insertion and Elimination: the doubly infinite Lie algebra of Feynman graphs}
\author{Alain Connes\thanks{IHES and Coll\`ege de France, connes@ihes.fr}
 $\;$and
Dirk Kreimer\thanks{Center for Mathematical Physics at Boston University and IHES, dkreimer@bu.edu}\\[5mm]
} \maketitle
\vspace{-20em}\begin{flushright}
{\small
BUCMP/02-01\\
hep-th/0201157}
\end{flushright}\vspace{16em}

\begin{abstract}
The Lie algebra of Feynman graphs gives rise to two natural
representations, acting as derivations on the commutative Hopf
algebra of Feynman graphs, by creating or eliminating subgraphs.
Insertions and eliminations do not commute, but rather establish a
larger Lie algebra of derivations which we here determine.
\end{abstract}
\section*{Introduction}

The algebraic structure of perturbative QFT
\cite{DK1,CK1,RHI,RHII} gives rise to commutative Hopf algebras
${\cal H}$ and corresponding Lie-algebras ${\cal L}$, with ${\cal
H}$ being the dual of the universal enveloping algebra of ${\cal
L}$. ${\cal L}$ can be represented by derivations of ${\cal H}$,
and two representations are most natural in this respect:
elimination or insertion of subgraphs.

Perturbation theory is indeed governed by a series over
one-particle irreducible graphs. It is then a straightforward
question how the basic operations of inserting or eliminating
subgraphs act. These are the basic operations which are needed to
construct the formal series over graphs which solve the
Dyson--Schwinger equations. We give an account of these actions
here as a further tool in the mathematician's toolkit for a
comprehensible description of QFT. We introduce these structures
by first considering the case of undecorated rooted trees. In that
case one is lead naturally to the two basic operations of grafting
and trimming using the relation between the Hopf algebras ${\cal
H}_{cm} $ and ${\cal H}_{rt}$ (\cite{CK1}). The Hopf algebra
${\cal H}_{cm} $ is neither commutative nor cocommutative but
admits a finite set of generators with simple relations. The basic
relation (\cite{CK1}) between a commutative subalgebra ${\cal
H}_{cm}^1$ of ${\cal H}_{cm} $ and the Hopf algebra ${\cal
H}_{rt}$ was obtained using the "natural growth operation" on
trees. By extending this "natural growth operation" to the
grafting of arbitrary trees we show how to enlarge ${\cal H}_{rt}$
to a Hopf algebra ${\cal H}_{rtt}$ whose relation to ${\cal
H}_{rt}$ is the same as the relation of ${\cal H}_{cm} $ with
${\cal H}_{cm}^1$. In particular it is neither commutative nor
cocommutative. We show that it is obtained as a "bicrossed
product" construction from a doubly infinite Lie algebra of rooted
trees, similar to the Lie algebra of formal vector fields. Since
most of the information is then contained in that Lie algebra,
which can be concretely described from grafting and trimming
operations, we then turn to Feynman graphs, and only discuss the
Lie algebra aspect in that case.

\section{Undecorated rooted trees}
\input psfig.sty
\def\d{\delta}
\def\be{\begin{equation}}
\def\ee{\end{equation}}
\def\Rb{{\mathbb{R}}}
\def\build#1_#2^#3{\mathrel{
\mathop{\kern 0pt#1}\limits_{#2}^{#3}}}


\subsection*{The Hopf algebras ${\cal H}_{cm} $ and ${\cal
H}_{rt}$} Let us first recall the constructions of the basic Hopf
algebras involved in \cite{CM} and \cite{CK1}, and compare their
properties.

\noindent As an algebra ${\cal H}_{cm}$ is noncommutative but
finitely generated.

\noindent It is
 generated by three elements $Y$, $X$, $\delta_1$. To describe the relations between these
three generators, one lets $\delta_n$, $n \geq 1$ be defined by
induction by,
\begin{equation}
 [X,\delta_n] = \delta_{n+1} \ \forall \, n \geq 1 \, , \label{eq0}
\end{equation}
then the presentation of the relations in ${\cal H}_{cm}$ is the
following,
\begin{equation}
[Y,X] = X , [Y,\delta_n] = n \, \delta_n , [\delta_n , \delta_m] =
0 \ \forall \, n , m \geq 1 ,
\end{equation}
The coproduct $\Delta$ in ${\cal H}_{cm}$ is defined by
\begin{equation}
\Delta \, Y = Y \otimes 1 + 1 \otimes Y \ , \ \Delta \, X = X
\otimes 1 + 1 \otimes X + \delta_1 \otimes Y \ , \ \Delta \,
\delta_1 = \delta_1 \otimes 1 + 1 \otimes \delta_1 \label{eq2}
\end{equation}
 and the equality,
\begin{equation}
\Delta (h_1 \, h_2) = \Delta h_1 \, \Delta h_2 \qquad \forall \,
h_j \in {\cal H}_T \, . \label{eq3}
\end{equation}
The Hopf algebra ${\cal H}_{cm}$ is neither commutative nor
cocommutative but is obtained in a simple manner from the
commutative subalgebra ${\cal H}_{cm}^1$ generated by the
$\delta_n$.
\medskip

\noindent {\bf Theorem.} (\cite{CM}) {\it Let $G_2$ be the group
of formal diffeomorphisms of the real line of the form  $\psi (x)
= \, x + \, o (x)$. For each $n$, let $\gamma_n$ be the functional
on $G_2$ defined by,}
\[
\gamma_n (\psi^{-1})= (\partial_x^n \log \psi' (x))_{x=0} \, .
\]
{\it The equality $ \Theta (\delta_n) =\, \gamma_n$ determines a
canonical isomorphism $\Theta$ of the Hopf algebra ${\cal
H}_{cm}^1$ with the Hopf algebra of coordinates on the group
$G_2$.} {\it The Hopf algebra ${\cal H}_{cm}$ is the bicrossed
product associated to the formal decomposition $G = G_1 \, G_2$
associated to the decomposition Lie $G=$ Lie $G_1 +$Lie $G_2$ of
formal vector fields in their affine part (Lie $G_1$) and
nilpotent part (Lie $G_2$).}

\medskip
\noindent The Hopf algebra ${\cal H}_{rt}$ of rooted trees is
commutative but not finitely generated.

\noindent Recall that a rooted tree $T$ is, by definition, a
finite, connected, simply connected, one dimensional simplicial
complex with a base point $* \in \Delta^0(T) = \{ \hbox{set of
vertices of}$ $T \}$. This base point is called the root. By the
degreee of the tree we mean
\begin{equation}
\vert T \vert = {\rm Card} \Delta^0(T) = \# \ \hbox{of vertices
of} \ T \, . \label{eq3.1}
\end{equation}
By a {\it simple} cut of a rooted tree $T$ we mean a subset $c
\subset \Delta^1(T)$ of the set of edges of $T$ such that,
\begin{equation}
\hbox{for any} \ x \in \Delta^0(T) \, \hbox{the path} \ (*,x) \
\hbox{only contains at most one element of} \ c \, . \label{eq3.3}
\end{equation}
Thus what is excluded is to have two cuts of the same path or
branch. Given a cut $c$ the new simplicial complex $T_c$ with
$\Delta^0(T_c) = \Delta^0(T)$ and
\begin{equation}
\Delta^1(T_c) = \Delta^1(T) \backslash c \, , \label{eq3.4}
\end{equation}
is no longer connected, unless $c = \emptyset$. We let $R_c (T)$
be the connected component of $*$ with the same base point and
call it the trunk. We endow each other connected component, called
a cut branch, with the base point coming from the cut. We obtain
in this way a set (with multiplicity) of rooted trees, which we
denote by $P_c (T)$. \noindent We let $\Sigma$ be the set of
rooted trees up to isomorphism, and let ${\cal H}_{rt}$ be the
polynomial commutative algebra generated by the symbols,
\begin{equation}
\delta_T \ , \ T \in \Sigma \, . \label{eq3.5}
\end{equation}
One defines a coproduct on ${\cal H}_{rt}$ by,
\begin{equation}
\Delta \, \delta_T = \delta_T \otimes 1 + 1 \otimes \delta_T +
\sum_c \left( \prod_{P_c (T)} \delta_{T_i} \right) \otimes
\delta_{R_c (T)} \, , \label{eq3.6a}
\end{equation}
where the last sum is over all non trivial simple cuts ($c \not=
\emptyset$) of $T$, while the product ${\displaystyle \prod_{P_c
(T)}}$ is over the cut branches.

\noindent Equivalently, one can write (\ref{eq3.6a}) as,
\begin{equation}
\Delta \, \delta_T = \delta_T \otimes 1 +  \sum_c \left(
\prod_{P_c (T)} \delta_{T_i} \right) \otimes \delta_{R_c (T)} \, ,
\label{eq3.7}
\end{equation}
where the last sum is over all simple cuts.

\noindent This defines $\Delta$ on generators and it extends
uniquely as an algebra homomorphism,
\begin{equation}
\Delta : {\cal H}_{rt} \rightarrow {\cal H}_{rt} \otimes {\cal
H}_{rt} \, . \label{eq3.8}
\end{equation}

\smallskip
\noindent The first basic relation between the Hopf algebras
${\cal H}_{cm}$ and ${\cal H}_{rt}$ is the Hopf algebra
homomorphism (\cite{CK1}) obtained using the "natural growth"
operator $N$ defined as the unique derivation of the commutative
algebra ${\cal H}_{rt}$ such that,
\begin{equation}
N \, \delta_T = \sum \, \delta_{T'} \label{eq3.51}
\end{equation}
where the trees $T'$ are obtained by adding one vertex and one
edge to $T$ in all possible ways without changing the base point.
It is clear that the sum (\ref{eq3.51}) contains $\vert T \vert$
terms.

\medskip
\noindent {\bf Theorem.} (\cite{CK1}) {\it The equality $ \Lambda
(\delta_n) =\, N^n(\delta_{*})$ determines a canonical
homomorphism $\Lambda$ of the Hopf algebra ${\cal H}_{cm}^1$ into
the Hopf algebra ${\cal H}_{rt}$ .}

\medskip
\noindent This theorem suggests, as we did in (\cite{CK1}) to
enlarge the Hopf algebra ${\cal H}_{rt}$ in the same way as ${\cal
H}_{cm}^1$ is naturally enlarged to ${\cal H}_{cm}$, by adjoining
the elements $Y, X$ implementing both the grading and the natural
growth operators. We shall now show that it is indeed possible to
do much more by extending the natural growth operator $N$ to the
grafting of arbitrary trees.

\subsection*{The derivations $N_T$ of ${\cal H}_{rt}$}

Let us first extend the construction of the natural
 growth operator $N$ to get operators $N_T$ labelled by arbitrary trees.

\noindent For a given rooted tree $T$ we consider the unique
derivation $N_T$ of ${\cal H}_{rt}$ such that, for any $t \in
\Sigma$, \be \label{eq1} N_T \, (\delta_t) = \sum_{v} \, \delta_{(
t \, \cup_{v} \, T)} \ee where in the summation, $v$ runs through
the vertices $v \in \Delta^0(t) $ and where the rooted tree $t'= t
\, \cup_{v} \, T$ is obtained as the union of $t$ and $T$, with
the root $*$ of $T$ identified with $v$. One has \be \Delta^1 (t
\, \cup_{v} \, T) = \Delta^1 (t) \, \cup \, \Delta^1 (T), \ee \be
{\rm root} (t \, \cup_{v} \, T) ={\rm root} (t), \ee

\noindent and the number of vertices of $(t \, \cup_{v} \, T)$ is,
\be \label{eq3b} \vert t' \vert = \vert t \vert + \vert T \vert -
1 \, . \ee When $T=*$ has one element we see that, \be \label{eq4}
N_* (\delta_t) = \vert t \vert \, \delta_t \ee thus we get the
derivation $Y$.

\noindent When $T = \srt$ is the rooted tree with one edge, we
just get the natural growth operation: $N_{\srt}=N$.

\noindent Since $N_T$ is extended as a derivation one has, \be
\label{eq5} N_T \left(\prod \, \delta_{t_i}\right) = \sum_1^n \,
\delta_{t_1} \ldots N (\delta_{t_k}) \ldots \delta_{t_n} \, . \ee
Let us now prove,

\medskip

\noindent {\bf Lemma.}
$$
\Delta (N_T (a)) = \left( N_T \otimes {\rm id} + {\rm id} \otimes
N_T +  \sum_c \, \prod_{P_c (T)} \, \delta_{t_j} \otimes N_{R_c
(T)} \right) \Delta (a) \, .
$$

\medskip

\noindent {\bf Proof.} Both sides of the equation are linear maps
from ${\cal H}_{rt}$ to ${\cal H}_{rt} \otimes {\cal H}_{rt}$
which satisfy the derivation rule, $\rho (ab) = \rho (a) \, \Delta
(b) + \Delta (a) \, \rho (b)$. Indeed ${\cal H}_{rt}$ is a
commutative algebra so that the multiplication by a product of
$\delta_{t_j} \otimes 1$ does not alter the derivation rule.
 Thus it is enough to check the lemma for $a = f
\delta_t, t \in \Sigma$.

\noindent Now, by definition of the coproduct,
$$
\Delta \, N_T (\delta_t) = N_T (\delta_t) \otimes 1 + 1 \otimes
N_T (\delta_t) + \sum_{v_0 , c} \, \prod \, \delta_{t'_j} \otimes
\delta_{R_c (t')}
$$
where $v_0$ varies in $ \Delta^0 (t)$ and the $c$ varies through
simple cuts of $t' = t \, \cup_{v_0} \, T$ .

\noindent Let us first consider the partial sum over pairs $(v_0 ,
c)$ with $v_0 \notin R_c (t')$ i.e. $v_0 \in \cup \, t'_i$.

\noindent This means that the segment $[* , v_0]$ is cut somewhere
and hence that $c \cap T = \emptyset$ since otherwise the cut
would not be simple.

\noindent We thus have $c \subset t$ so that we can view $c$ as a
cut of $t$. Thus $R_c (t')=R_c (t)$. Also $v_0 \in \cup \, t'_i$
and the sum over $v_0$  decomposes as a sum over $i$ and yields
for each $i$ the value \be \prod \, \delta_{t'_j}= N_T
(\delta_{t_i}) \prod_{j \not= i} \delta_{t_i} \ee Thus, since
$N_T$ is a derivation, the partial sum gives

\be \label{eq6} \sum_{c \, ({\rm cut \, of} \, t)} N_T \left(
\prod_{P_c (t)} \delta_{t_i} \right) \otimes \delta_{R_c (t)} \, .
\ee Now this equals $N_T \left( \build\sum_{c}^{} \,
\build\prod_{P_c (t)}^{} \delta_{t_i} \right) \otimes \delta_{R_c
(t)}$ and we can group this sum with $N_T (\delta_t) \otimes 1$,
using $N_T (1) = 0$ to get, \be \label{eq7} (N_T \otimes {\rm id})
\, \Delta (\delta_t) \, . \ee which is the first term in the right
hand side of the equation of the lemma.

\noindent We then consider the partial sum over pairs $(v_0 , c)$
with $v_0 \in R_c (t')$ and $c \cap T = \emptyset$.
 Then $c$ is a
cut of $t$ as above, while $v_0$ now varies among the vertices of
$R_c (t)$. One has $t'_i=t_i$ and $R_c (t')=R_c (t)\cup_{v_0}T$.
Thus the sum over $v_0$ replaces  $\delta_{ R_c (t)}$ by $N_T
(\delta_{R_c (t)})$ without touching the $\delta_{t_i}$. We can
group this with $1 \otimes N_T (\delta_t)$ and get, \be
\label{eq8} ({\rm id} \otimes N_T) \, \Delta (\delta_t) \, . \ee
which is the second term in the right hand side of the equation of
the lemma.

\noindent We are now left only with the partial sum over pairs
$(v_0 , c)$ such that $c \, \cap \, T \ne \emptyset$ (in which
case $v_0 \in R_c (t')$). Let us then fix the nonempty simple cut
of $T$, \be \label{eq9} c' = c \cap \Delta^1 (T) \, , \ee and show
that the corresponding partial sum is equal to, \be \label{eq11}
\prod_{t_j \in P_{c'} (T)}  \, \delta_{t_j} \otimes N_{R_{c'} (T)}
(\Delta \, \delta_t) \, . \ee Since $\Delta^1 (t') = \Delta^1 (t)
\, \cup \, \Delta^1 (T)$, one has $c=c_1 \, \cup \, c'$ where
$c_1$ now varies among (possibly empty) simple cuts of $t$.
Moreover $v_0$ now varies in $ R_{c_1} (t)$.

\noindent To each $\varepsilon \in c=c_1 \, \cup \, c'$ there is a
corresponding fallen branch $t_{\varepsilon}$. For $\varepsilon
\in c_1$ it is a fallen branch of $t$ for $c_1$
 while for $\varepsilon \in c'$ it is a fallen
branch of $T$ for $c'$. Thus the product of fallen branches is \be
\label{eq10} \prod_{t_i \in P_{c_1} (t)} \, \delta_{t_i} \
\prod_{t_j \in P_{c'} (T)} \, \delta_{t_j} \, . \ee One has
\begin{equation}
\Delta \, \delta_t = \delta_t \otimes 1 +  \sum_{c_1} \left(
\prod_{P_{c_1} (t)} \delta_{t_i} \right) \otimes \delta_{R_{c_1}
(t)} \, , \label{eq3.6}
\end{equation}
where ${c_1}$ varies among (possibly empty) simple cuts of $t$.
Let $P = \build\prod_{t_j \in P_{c'} (T)}^{} \, \delta_{t_j}$ and
let us look at the terms in, \be \label{eq11b} (P \otimes
N_{R_{c'} (T)}) (\Delta \, \delta_t) \, . \ee The term $\delta_t
\otimes 1$ does not contribute since $N(1) = 0$. When we apply $P
\otimes N_{R_{c'} (T)}$ to the term $\build\prod_{P_{c_1} (t)}^{}
\, \delta_{t_j} \otimes \delta_{ R_{c_1} (t)}$  in $\Delta \,
\delta_t$, we get \be \label{eq13} \sum_{v_0} P \prod_{P_{c_1}
(t)} \, \delta_{t_j} \otimes  \delta_{R_{c_1} (t) \, \cup_{v_0} \,
R_{c'} (t')} \, . \ee where $v_0$ varies in $R_{c_1} (t)$.

\noindent With $t' = t \, \cup_{v_0} \, T$, one has $R_{c_1} (t)
\, \cup_{v_0} \, R_{c'} (T) = R_c (t')$, for $c = c_1 \cup c'$.
Thus we get the
 corresponding term of $\Delta (N_T
(\delta_t))$, namely, \be \label{eq12} P \prod_{t_j \in P_{c_1}
(t)} \, \delta_{t_j} \otimes \delta_{R_c (t')} \, . \ee
 Taking the sum over pairs
$(v_0 , c_1)$ such that $v_0 \in R_{c_1} (t)$ yields the required
equality and completes the proof of the lemma.~$\|$

\medskip

\noindent It is then natural to enlarge the Hopf algebra ${\cal
H}_{rt}$ by introducing new generators $X_T, T \in \Sigma$ such
that, \be \label{eq14} [X_T , \delta_t ] = N_T (\delta_t) \ee and
with coproduct rule given by, \be \label{eq15} \Delta \, X_T = X_T
\otimes 1 + 1 \otimes X_T + \sum_c \prod_{P_c (T)} \delta_{t_j}
\otimes X_{R_c (T)} \, . \ee This coproduct is superficially
similar to (\ref{eq3.6}), but the right hand side now involves
both the $\delta$'s and the $X$'s. In order to complete the
presentation of the extended Hopf algebra ${\cal H}_{rtt}$, we
need to compute the Lie bracket of the derivations $N_T$. This is
straightforward and given by
\medskip

\noindent {\bf Lemma.} \be [N_{T_1} , N_{T_2}] = \sum_{v_2 \in
\Delta^0 (T_2)} N_{T_2 \cup_{v_2} T_1} - \sum_{v_1 \in \Delta^0
(T_1)} N_{T_1 \cup_{v_1} T_2} \, . \ee

\medskip

\noindent  We are dealing with derivations of ${\cal H}_{rt}$ and
it is thus enough to consider the action of both sides on
$\delta_t$. One has,
$$
N_{T_1} (N_{T_2} (\delta_t)) = \sum_{v_0 \in \Delta^0 (t)} N_{T_1}
(\delta_{t \cup_{v_0} T_2}) = \sum_{v_0 \in \Delta^0 (t)} \
\sum_{v_1 \in \Delta^0 (t \cup_{v_0} T_2)} \delta_{(t \cup_{v_0}
T_2) \cup_{v_1} T_1}
$$
$$
= \sum_{v_0 \in \Delta^0 (t)} \ \sum_{v_1 \in \Delta^0 (T_2)}
\delta_{t \cup_{v_0} (T_2 \cup_{v_1} T_1)} + \sum_{{v_0 , v_1 \in
\Delta^0 (t) \atop v_0 \ne v_1}} \delta_{t \cup_{v_0} T_1
\cup_{v_1} T_2} \, .
$$
The last term is symmetric in $T_1 , T_2$ and thus does not
contribute to the commutator which is thus given by the formula of
the lemma.

\medskip
\noindent We can thus complete the presentation of the Hopf
algebra ${\cal H}_{rtt}$ by the rule, \be \label{eq18} [X_{T_1} ,
X_{T_2}] = \sum_{v_2 \in \Delta^0 (T_2)} X_{T_2 \cup_{v_2} T_1} -
\sum_{v_1 \in \Delta^0 (T_1)} X_{T_1 \cup_{v_1} T_2} \, . \ee and
define ${\cal H}_{rtt}$ as the envelopping algebra of the Lie
algebra which is the linear span of the $X_T, \delta_t, T,t \in
\Sigma$, with bracket given by (\ref{eq18}), (\ref{eq14}) and the
commutativity of the $\delta$'s. We define a coproduct on ${\cal
H}_{rtt}$ by  (\ref{eq3.6}) and (\ref{eq15}). We thus get,
\medskip

\noindent {\bf Theorem. } {\it Endowed with the above structure
${\cal H}_{rtt}$ is a Hopf algebra.} {\it The equalities $ \Lambda
(\delta_n) =\, N^n(\delta_{*})$, $ \Lambda (Y) = \, X_{*}$, $
\Lambda (X) =\, (X_{\srt})$ determine a canonical homomorphism
$\Lambda$ of the Hopf algebra ${\cal H}_{cm}$ in the Hopf algebra
${\cal H}_{rtt}$ .}

\medskip

\noindent The best way to comprehend the Hopf algebra structure of
${\cal H}_{rtt}$ is to consider the natural action of ${\cal
H}_{rtt}$ as an algebra on the dual of ${\cal H}_{rt}$, obtained
by transposition. The compatibility of the algebra structures
dictates the Hopf algebra structure, by transposing multiplication
to comultiplication. Combining the basic Hopf algebra identity,
$m(S \otimes Id) \Delta = \epsilon$ with equation (31) yields the
following explicit formula for the antipode $S(X_T)$, $T \in
\Sigma$, \be \label{xxxx}
 S( X_T) = - X_T - \sum_c \prod_{P_c (T)} S(\delta_{t_j})
 X_{R_c (T)} \, , \ee
using the known formula for $S(\delta_{t_j})$ in the subalgebra
${\cal H}_{rt}$.

\noindent The reader should note that $S^2\not= 1$ for the
antipode $S$  in ${\cal H}_{rtt}$, as this algebra is neither
commutative nor cocommutative, comparable to the situation in
${\cal H}_{cm}$. Indeed, we now have a large supply of natural
growth operators in generalization of that situation.

Let $ \overline{\Delta}(X_T)=\Delta(X_T)-X_T\otimes 1-1\otimes
X_T.$ For the multiple application of that subtracted coproduct,
we can still uniquely write
$$\overline{\Delta}^n(X_T)=X_{T^\prime }\otimes\cdots\otimes X_{T^{\prime\ldots\prime  }},\;\mbox{\tiny $n+1$ $\prime$ s}.$$
It is obvious that the Hopf algebra endomorphism $S^2$ fulfills
$S^2(\delta_t)=\delta_t$, while for the generators $X_T$ we have
\medskip

\noindent {\bf Proposition.} {\it
$$S^2(X_T)=X_T+N_{T^{\prime\prime}}(\delta_{T^\prime})+S(\delta_{T^{\prime\prime}})N_{T^{\prime\prime\prime}}(\delta_{T^\prime}).$$}
\medskip

\noindent Proof: In the above notation,
$S(X_T)=-X_T-S(\delta_{T^\prime})X_{T^{\prime\prime}}$, and also
$S(\delta_{T^\prime})X_{T^{\prime\prime}}=\delta_{T^\prime}S(X_{T^{\prime\prime}})$.
Thus
\begin{eqnarray*}
S^2(X_T) & = & S[-X_T-S(\delta_{T^\prime})X_{T^{\prime\prime}}]\\
 & = &
 X_T+S(\delta_{T^\prime})X_{T^{\prime\prime}}-S(X_{T^{\prime\prime}})\delta_{T^\prime}\\
  & = &
  X_T+S(\delta_{T^\prime})X_{T^{\prime\prime}}-\delta_{T^\prime}S(X_{T^{\prime\prime}})-[S(X_{T^{\prime\prime}}),\delta_{T^\prime}]\\
  & = & X_T -[S(X_{T^{\prime\prime}}),\delta_{T^\prime}]\\
    & = &
    X_T+N_{T^{\prime\prime}}(\delta_{T^\prime})+S(\delta_{T^{\prime\prime}})N_{T^{\prime\prime\prime}}(\delta_{T^\prime}),
\end{eqnarray*}
using $\overline{\Delta}^2$.~$\|$
\medskip

\noindent It is of course desirable to extend to the Hopf algebra
${\cal H}_{rtt}$ the description of ${\cal H}_{cm}$ as a bicrossed
product associated to the decomposition Lie $G=$ Lie $G_1 +$Lie
$G_2$ of the Lie algebra of formal vector fields. \noindent Our
next task will be to describe the Lie algebra ${\cal L}$
 that will play the role
of the Lie algebra of formal vector fields.

\noindent As a preliminary remark, let us relate the Lie algebra
structure ${\cal L}_1$ on the $X_T$ given by (\ref{eq18}) to an
operad ${\cal P}$. This insertion operad \cite{Jim} underlies the
pre-Lie structure, whose antisymmetrization is the Lie bracket
(\ref{eq18}). The operad is obtained by considering as elements of
${\cal P} (n)$ a pair of rooted tree $t$ and a bijection, \be
\label{eq19} \sigma : \{ 1 , \ldots , n \} \rightarrow \Delta^0
(t) \, . \ee We then define $t \circ_i t'$ as $t \, \cup_{\sigma
(i)} \, t'$, for $i \in \{ 1 , \ldots , n \}$ and where the new
bijection is obtained by shifting the labels of the vertices
$\sigma (i+1) \ldots \sigma (n)$ to $i+n' , \ldots , n+n'-1$ as
well as the labels of the vertices $\sigma' (1) \ldots \sigma'
(n')$ to $i , i+1 , \ldots , i+n'-1$.

$$
\hbox{ \psfig{figure=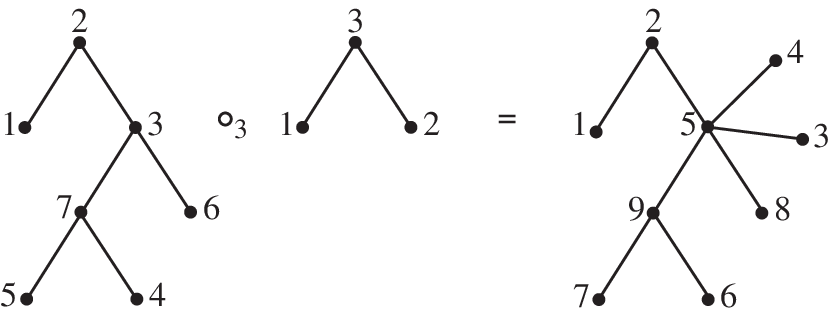} }
$$

\noindent One has a natural action of $S_n$ the group of
permutations of $\{ 1 , \ldots , n \}$ which replaces $\sigma$ by
$\sigma \circ \pi^{-1}$, i.e. replaces the labelling $\sigma^{-1}$
of the vertices by $\pi \circ \sigma^{-1}$. One checks that, \be
\label{eq20} t^{\pi} \circ_{\pi (i)} t'^{\rho} = (t \circ_i
t')^{\alpha} \ee where $\alpha$ is obtained from the permutations
$\pi$ of $\{ 1 , \ldots , n \}$, $\rho$ of $\{ 1 , \ldots , n' \}$
and $i \in \{ 1 , \ldots , m\}$ by acting by $\rho$ in $\{
i,i+1,\ldots ,i+n'-1\}$ and by $\pi$ after collapsing the above
interval to $\{ i \}$.

\noindent One also checks the following two equalities for
$\lambda \in {\cal P} (\ell)$, $\mu \in {\cal P} (m)$, $\nu \in
{\cal P} (n)$ \be \label{eq21} (\lambda \circ_i \mu) \circ_{j+m-1}
\nu = (\lambda \circ_j \nu) \circ_i \mu \qquad 1 \leq i < j \leq
\ell \ee \be \label{eq22} (\lambda \circ_i \mu) \circ_{i-1+j} \nu
= \lambda \circ_i (\mu \circ_j \nu) \qquad 1 \leq i \leq \ell \, ,
\ 1 \leq j \leq m \, . \ee The first is the independence of two
graftings at two distinct vertices, and the second is a kind of
associativity of grafting.

\subsection*{The Lie algebra ${\cal L}$}

\noindent We shall now describe the Lie algebra ${\cal L}= {\cal
L}_1 + {\cal L}_2$ playing the role of the Lie algebra of formal
vector fields, in the case of rooted trees, i.e.~baring the same
relation with ${\cal H}_{rtt}$ as the Lie algebra of formal vector
fields does with ${\cal H}_{cm}$. We already know the Lie
subalgebra $ {\cal L}_1$ of the $X_T$'s. The Lie algebra ${\cal
L}_2$ is the Lie algebra of primitive elements in the dual of
${\cal H}_{rt}$. In order to obtain ${\cal L}$ we consider the
natural actions of both $ {\cal L}_1$ and $ {\cal L}_2$ as
derivations of the commutative algebra ${\cal H}_{rt}$. We already
saw the action $N$ of $ {\cal L}_1$. The action of $ {\cal L}_2$
is the canonical action of the Lie algebra of primitive elements
of the dual of ${\cal H}_{rt}$ on the commutative algebra ${\cal
H}_{rt}$. It is given by the following derivations $M_T$ of ${\cal
H}_{rt}$, \be \label{eq2.2} M_T (a) = \langle Z_T \otimes {\rm id}
, \Delta (a) \rangle \qquad \forall \, a \in {\cal H} \, , \ee
where, for $T \in \Sigma$, $Z_T$ is the primitive element of the
dual ${\cal H}_{rt}^*$ given by the linear form on ${\cal H}_{rt}$
which vanishes on any monomial $\delta_1 \, \delta_2 \,..... \,
\delta_n $ except for $\delta_T $, with \be \langle Z_T, \delta_T
\rangle = 1. \label{eq2.200} \ee

 \noindent One has $Z_T (ab) = Z_T (a) \, \varepsilon (b) + \varepsilon
(b) \, Z_T (b)$ so that by construction $M_T$ is a derivation of
${\cal H}$.

 \noindent The Lie bracket of the $Z_T$'s is given by the Lie
algebra of rooted trees, i.e. \be \label{eq2.1} [Z_{T_1} ,
Z_{T_2}] = \sum (n (T_1 , T_2 ; T) - n (T_2 , T_1 ; T)) \, Z_T \,
. \ee where $n (T_1 , T_2 ; T)$ is the number of cuts $c$ of $T$
of cardinality one $(\vert c \vert = 1)$ such that $P_c(T) = T_1,
\, R_c (T)=T_2$.

\noindent For $a = \delta_t$ we get, \be \label{eq2.3} M_T
(\delta_t) = \sum_{{\vert c \vert = 1 \atop P_c(t) = {T}}}
\delta_{ R_c (t)} \qquad \hbox{if} \ t \ne T \ee and, \be
\label{eq2.4} M_T (\delta_T) = 1 \, . \ee Thus $M_T (\delta_t) =
0$ unless $t=T$ or $t$ admits an edge $\varepsilon \in \Delta^1
(t)$ such that $P_{\varepsilon} (t) = T$.

$$
\hbox{ \psfig{figure=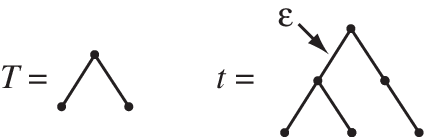} }
$$

\noindent By construction $M$ is a representation of the Lie
algebra
 $ {\cal L}_2$ in the Lie algebra of derivations $D$ of ${\cal H}_{rt}$
 which preserve the linear span ${\cal
D}= \{ \Sigma \, \lambda_T \, \delta_T + \lambda_1 1 \}$, \be
\label{eq2.5} D \,({\cal D}) \subset {\cal D}. \ee Similarly the
representation $N$ of  $ {\cal L}_1$ is given by derivations
fulfilling (\ref{eq2.5}). In order to show that $ {\cal L}= {\cal
L}_1 + {\cal L}_2$ is a Lie algebra, let us now compute the
commutator \be \label{eq2.6} M_{T_1} \, N_{T_2} - N_{T_2} \,
M_{T_1} \, . \ee Let us first consider the case where $T_1$ and
$T_2$ are not comparable, i. e. we assume that $T_1 \ne P_c (T_2)$
for all cuts $c$, $\vert c \vert = 1$ of $T_2$ and that $T_1 \ne
t_1 \, \cup_v \, T_2$ for any tree $t_1$ and vertex $v \in
\Delta^0 (t_1)$ . Let us show that in that case $M_{T_1}$ and
$N_{T_2}$ actually commute. The nonzero terms in $M_{T_1} \,
N_{T_2} (t)$ are given by $\delta_{P_{\varepsilon}(t \, \cup_{v_0}
\, T_2)}$ for a vertex $v_0 \in \Delta^0 (t)$ and an edge
$\varepsilon \in \Delta^1 (t \, \cup_{v_0} \, T_2)$ such that
$P_{\varepsilon}(t \, \cup_{v_0} \, T_2) = T_1$.

\noindent Now $\Delta^1 (t \, \cup_{v_0} \, T_2) = \Delta^1 (t) \,
\cup \, \Delta^1 (T_2)$, and if $\varepsilon \in \Delta^1 (T_2)$
would yield a nonzero term, then $T_1$ would appear as
$P_{\varepsilon} (T_2)$. Thus $\varepsilon \in \Delta^1 (t)$.

\noindent Next if $v_0 \notin R_{\varepsilon} (t)$ then $v_0 \in
P_{\varepsilon} (t)$ and $P_{\varepsilon} (t \, \cup_{v_0} \, T_2)
= P_{\varepsilon} (t) \, \cup_{v_0} \, T_2$. But by hypothesis
this cannot be $T_1$ so we get 0.

\noindent The only remaining case is $v_0 \in R_{\varepsilon} (t)$
so that $P_{\varepsilon} (t \, \cup_{v_0} \, T_2) =
P_{\varepsilon} (t)$ while $R_{\varepsilon} (t \, \cup_{v_0} \,
T_2) = R_{\varepsilon} (t) \, \cup_{v_0} \, T_2$, thus we get,
\begin{equation}
M_{T_1} \, N_{T_2} (t) = \sum_{{v_0 \in \Delta^0 (t) , \varepsilon
\in \Delta^1 (t) \atop v_0 \in R_{\varepsilon} (t) ,
P_{\varepsilon} (t) = T_1}} \ \delta_{R_{\varepsilon} (t) \,
\cup_{v_0} \, T_2} \, .
\end{equation}
But we have
\begin{equation}
M_{T_1} (t) = \sum_{\varepsilon \in \Delta^1 (t) , P_{\varepsilon}
(t) = T_1}\  \delta_{R_{\varepsilon} (t)}
\end{equation}
and
\begin{equation}
N_{T_2} (M_{T_1} (t)) = \sum_{{\varepsilon \in \Delta^1 (t) , v_0
\in R_{\varepsilon} (t) \atop P_{\varepsilon} (t) = T_1}} \
\delta_{R_{\varepsilon} (t) \, \cup_{v_0} \, T_2 }\, .
\end{equation}
Thus we see that if $T_1$ and $T_2$ are not comparable we get \be
\label{eq2.7} M_{T_1} \, N_{T_2} = N_{T_2} \, M_{T_1} \, . \ee In
general, given $t, T_1, T_2 \in \Sigma$ we define the
 integers $N (t , T_2 ; T_1)$  and $M (T_1 , T_2 ; t)$ by,
\be \label{eq2.9}
 N (t , T_2 ; T_1) = \langle N_{T_2} (\delta_t) , Z_{T_1} \rangle
\ee and \be \label{eq2.11} M (T_1 , T_2 ; t) =\langle M_{T_1}
(\delta_{T_2}) , Z_t \rangle. \ee By construction $N (t , T_2 ;
T_1)$
 is the number of times $T_1$ occurs as $t \, \cup_v \,
T_2$ while  $M (T_1 , T_2 ; t)$ is the number of times $T_1$
occurs as $P_c(T_2)$ with $\vert c \vert = 1$ and $R_c(T_2)=t$. We
then get,
\medskip

\noindent {\bf Lemma.} \be \label{eq2.10} [M_{T_1} , N_{T_2}] =
\sum_{t} N (t , T_2 ; T_1) \, M_{t} \, + \sum_{t} M (T_1 , T_2 ;
t) \, N_{t} \, . \ee

\medskip

\noindent First assume $\vert T_1 \vert \geq \vert T_2 \vert$ so
that $T_1$ cannot be a $P_c (T_2)$, for $\vert c \vert = 1$ and $M
(T_1 , T_2 ; t)=0$. Then the same computation of $[M_{T_1} \, ,
N_{T_2}](\delta_t) $ as above gives the sum of the
$\delta_{R_{\varepsilon} (t)}$ such that $T_1$
 occurs as a $P_{\varepsilon} (t)
\, \cup_{v_0} \, T_2$. Fixing then $t_1=P_{\varepsilon} (t)$ we
see that we obtain the sum of the $M_{t_1}$ with multiplicity
given by the number of solutions of \be \label{eq2.8} t_1 \,
\cup_v \, T_2 = T_1 \, . \ee

\noindent Next assume that $\vert T_1 \vert < \vert T_2 \vert$ so
that $T_1$ can occur as $P_c (T_2)$, $\vert c \vert = 1$, but
cannot occur as $t_1 \, \cup_v \, T_2$, so that $ N (t , T_2 ;
T_1)=0$. Then in the above computation of $[M_{T_1} \, ,
N_{T_2}](\delta_t) $ the case $v_0 \in P_{\varepsilon} (t)$ above
only gives 0 and the only nonzero contribution comes when
$\varepsilon \in \Delta^1 (T_2)$. One then has $R_{\varepsilon} (t
\, \cup_{v_0} \, T_2) = t \, \cup_{v_0} \, R_{\varepsilon} (T_2)$
and $P_{\varepsilon} (t \, \cup_{v_0} \, T_2) = P_{\varepsilon}
(T_2)$ which must be $ T_1$ to yield a non zero result. Thus we
obtain the sum of the $\delta_{ R_{\varepsilon} (t \, \cup_{v_0}
\, T_2)}$ where $P_{\varepsilon} (T_2)=T_1$. This equals the sum
of the $\delta_{t \, \cup_{v_0} \, R_{\varepsilon} (T_2)}$ and
hence, letting $t_2 =R_{\varepsilon} (T_2)$ the sum of the $ M
(T_1 , T_2 ; t_2) \, N_{t_2}(\delta_t)$. We need to take care of
(\ref{eq2.4}), i.e. to consider the case where $M_{T_1}$ is
applied to some $t \, \cup_{v_0} \, T_2 = T_1$ which only occurs
when $\vert T_1 \vert \geq \vert T_2 \vert$. For each such term
one takes $c = \emptyset$ so the above discussion does not apply,
but one can check that the additional contribution to both sides
of (\ref{eq2.10}) do agree when evaluated on $t$ fulfilling
(\ref{eq2.8}) for some $v \in \Delta^0 (t)$.

\medskip
\noindent We can now define the full Lie algebra ${\cal L}$ of
rooted trees by introducing new generators of the form, $Z_{-t}$
where $t$ is a rooted tree, and extending the Lie bracket
(\ref{eq2.1}) based on the above lemma. We associate $Z_{-T}$ with
$-N_T$ and $Z_T$ with $M_T$ and work out the Lie brackets so that
we get a representation.
 In particular the
elements $Z_0$, $Z_{-1}$ now become, \be \label{eq2.13} Z_0 =
Z_{-*} , Z_{-1} = Z_{-T} \, . \ee
 We
use the $-$ sign, $-N_T$ to get that the commutator with $Z_{-*}$
does give the grading of the Lie algebra. Indeed if we apply
(\ref{eq2.10}) for $T_2 = *$ we get \be \label{eq2.14} [-N_* ,
M_T] = \vert T \vert \, M_T \, , \ee while one has, \be
\label{eq2.15} [-N_* , N_T] = (1 - \vert T \vert) \, N_T \, . \ee
\medskip

\noindent {\bf Theorem. } {\it ${\cal L}={\cal L}_1 +{\cal L}_2$
is a Lie algebra.}\\
{\it The Hopf algebra ${\cal H}_{rtt}$ is the bicrossed product
associated to the
 decomposition ${\cal L}={\cal L}_1 +{\cal L}_2$.}

\medskip

\noindent As a final remark, note that the Lie subalgebra ${\cal
L}_2$ generated by the $Z_T$ is naturally isomorphic to a
subalgebra of ${\cal L}_1$ generated by the $Z_{-T}$. Indeed one
lets $*T$ be the new rooted tree given by, \be \label{eq2.16}
\hbox{ \psfig{figure=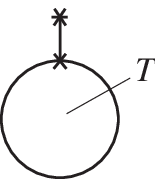} } \ee Then the following map is an
inclusion ${\cal L}_2 \subset {\cal L}_1$, \be \label{eq2.17} Z_T
\rightarrow \frac{1}{S_T} \ Z_{- (*T)} \, . \ee By (\ref{eq2.1})
we see that this is a Lie algebra homomorphism since the grafting
at $*$ gives a symmetric result, which drops out of the bracket:
$$
\hbox{ \psfig{figure=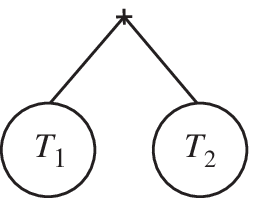} }.
$$
\section{Graphs}
\subsection*{Formal Definitions} We only consider graphs without
self-loops: no edge starts and ends in the same vertex. We allow
for multiple edges though: two vertices might be connected by more
than one edge.

\noindent First, we define $n$-particle irreducible ($n$-PI)
graphs.\\[2mm] {\bf Definition.} {\it  A $n$-particle irreducible graph
$\Gamma$ is graph such that upon removal of any set of $n$ of its
edges it is still connected. Its set of edges is denoted by
$\Gamma^{[1]}$ and its set of vertices is denoted by
$\Gamma^{[0]}$, edges and vertices can be of various different
type.\\[2mm]} The type of an edge is often indicated by the way we
draw it: (un-)oriented straight lines, curly lines, dashed lines
and so on. These types of edges, often called propagators in
physicists parlance, are chosen in accordance to Lorentz covariant
wave equations: the propagator as the analytic expression assigned
to an edge is an inverse wave operator with boundary conditions
typically chosen in accordance with causality.

\noindent The types of vertices are determined by the types of
edges to which they are attached:\\[2mm] {\bf Definition.} {\it  For
any vertex $v$ $\in \Gamma^{[0]}$ we call the set
$f_v:=\{f\in\Gamma^{[1]}\mid v\cap f\not=\emptyset \} $ its
type.\\[2mm]} Note that $f_v$ is a set of edges.

\noindent Of particular importance are the 1PI graphs. They
decompose into disjoint graphs upon removal of an edge. Note that
any $n$-PI graphs is also $(n-1)$-PI, $ \forall n\geq 1$. A graph
which is not 1-PI is called reducible. Also, any connected graph
is considered as 0-PI.

\noindent A further notion needed is the one of external and
internal edges.\\[2mm] {\bf Definition.} {\it  An edge $f$ $\in
\Gamma^{[1]}$ is internal, if $\{v_f\}:=f\cap \Gamma^{[0]}$ is a
set of two elements.\\[2mm]} So, internal edges connect two vertices
of the graph $\Gamma$.\\[2mm] {\bf Definition.} {\it  An edge $f$ $\in
\Gamma^{[1]}$ is external, if $f\cap \Gamma^{[0]}$ is a set of one
element.\\[2mm]} As we exclude self-loops, this means that an external
edge has an open end. Thus external edges are associated with a
single vertex of the graph. These edges correspond to external
particles interacting in the way prescribed by the graph. There
are obvious gluing operations combining 1PI graphs into reducible
graphs, by identifying two open ends of edges of the same type
originating from different 1PI graphs. We will make no use of
reducible graphs here but note that the Hopf and Lie algebra
structures could be set up in this context as well.

\noindent $\Gamma^{[1]}$ obviously decomposes into the set of
internal edges and the set of external edges of a graph $\Gamma$,
$$\Gamma^{[1]}=\Gamma^{[1]}_{\rm ext}\cup\Gamma^{[1]}_{\rm int}.$$

\noindent We now turn to the possibilities of inserting graphs
into each other. Our first requirement is to establish bijections
between sets of edges so that we can define gluing operations.\\[2mm]
{\bf Definition.} {\it  We call two sets of edges $I_1,I_2$
compatible, $I_1\sim I_2$, iff they contain the same number of
edges, of the same type.\\[2mm]} Compatibility is an equivalence
relation. We will utilize it to glue graphs into each other. To
compare vertices, we look at the adjacent edges:\\[2mm] {\bf
Definition.} {\it  Two vertices $v_1,v_2$ are of the same type, if
$f_{v_1}$ is compatible with $f_{v_2}$.\\[2mm]} Quite often, we will
shrink a graph to a point. The only useful information still
available after that process is about its set of external edges:\\[2mm]
{\bf Definition.} {\it  We define ${\bf res}(\Gamma)$ to be the
result of identifying $\Gamma^{[0]}\cup\Gamma^{[1]}_{\rm int}$
with a point in $\Gamma$.\\[2mm]} An example is $$\res.$$ Note that
${\bf res}(\Gamma)^{[1]}\equiv {\bf res}(\Gamma)^{[1]}_{\rm
ext}\sim \Gamma^{[1]}_{\rm ext}$. By construction all graphs which
have compatible sets of external edges have the same residue.

\noindent If the set $\Gamma^{[1]}_{\rm ext}$ is empty, we call
$\Gamma$ a vacuumgraph, if it contains a single element we call
the graph a tadpole graph. Vacuum graphs and tadpole graphs will
be discarded in most of what follows. If this set contains two
elements, we call $\Gamma$ a self-energy graph, if it contains
more than two elements, we call it an interaction graph. Further
we restrict ourselves to graphs which have vertices such that the
cardinality of their types is $\geq 2$. If needed, for example in
the presence of external fields, this can be relaxed.

\noindent A further important notion is the gluing of graphs into
each other. It is the opposite of the shrinking of a graph to its
residue. While in that process, a graph is reduced to a vertex of
a specified type, we can replace any vertex $v\in \Gamma^{[0]}$ of
type $f_v$ by a graph $\gamma$, as long as $f_v\sim
\gamma^{[1]}_{\rm ext}$ - a vertex will be replaced by a graph
which has external edges compatible with its type.

\noindent To specify such a gluing of $\gamma$ into $\Gamma$ we
first have to choose an internal vertex $v$ where we wish to glue.
If the type of $v$ is incompatible with $\gamma^{[1]}_{\rm ext}$,
we define the result to vanish. If the two sets of edges are
compatible, we will have in general to choose a bijection between
the two sets of edges. Summing over all places and bijections
defines an operation $\Gamma\star\gamma$ which sums over all ways
of inserting $\gamma$ into $\Gamma$. We impose a normalization
such that topologically different graphs are generated with unit
multiplicity. The following picture illustrates this process.
$$\picf .$$
{\bf Proposition.} {\it  This gluing operation is pre-Lie.\\[2mm]}
Proof: It suffices to show that for 1PI graphs $\Gamma_i$,
$i=1,2,3$, we have
$$\Gamma_1\star(\Gamma_2\star\Gamma_3)-(\Gamma_1\star\Gamma_2)\star\Gamma_3
=\Gamma_1\star(\Gamma_3\star\Gamma_2)-(\Gamma_1\star\Gamma_3)\star\Gamma_2
.$$ This is elementary using that both sides reduce to the sum
over all ways of gluing $\Gamma_2$ and $\Gamma_3$ simultaneously
into $\Gamma_1$ at disjoint places.~$\|$\\[5mm]
Note that this pre-Lie operation can be extended to the insertion
at internal edges (self-energies). Furthermore, external
structures \cite{RHI} can be incorporated easily, using coloured
types of vertices.

\noindent Choices of types of lines and vertices are typically
dictated by a chosen QFT, where, in particular, one often only
consider superficially divergent graphs. External structures
reflect their powercounting degree of divergence.

\noindent We let ${\cal L}_{FG}$ be any such chosen Lie-algebra
generated from this pre-Lie product, and ${\cal H}_{FG}$ be the
commutative Hopf algebra which we obtain as the dual of the
universal enveloping algebra of ${\cal L}_{FG}$.

\subsection*{Derivations on the Hopf algebra} We have the
decomposition of ${\cal H}_{FG}$ by the bidegree ${\cal
H}_{FG}=\oplus_{k=0}^\infty {\cal H}_{FG}^{[k]}$, reduced to
scalars $\in {\cal H}_{FG}^{[0]}$ by the counit. The linear basis
of ${\cal H}_{FG}$ is denoted by ${\cal H}_{FG,L}$. It is spanned
by generators $\delta_\Gamma$, where $\Gamma$ is a 1PI graph.
Elements of ${\cal H}_{FG}$ are polynomials in these commutative
variables.

\noindent We write $Z_\Gamma$ for the dual basis of the universal
enveloping algebra with pairing $$ \langle
Z_\Gamma,\delta_{\Gamma^\prime}\rangle=\delta^K_{\Gamma,\Gamma^\prime},$$
where on the rhs we have the Kronecker $\delta^K$, and extend the
pairing by means of the coproduct $$\langle
Z_{\Gamma_1}Z_{\Gamma_2},X\rangle =\langle Z_{\Gamma_1}\otimes
Z_{\Gamma_2},\Delta(X) \rangle.$$

\noindent For $X=\sum_i c_i\Gamma_i$, we extend by linearity so
that $\delta_X=\sum_i c_i\delta_{\Gamma_i}$, and similarly for
$Z_X$.

\noindent Quite often, we want to refer to the graph(s) which
index an element in  ${\cal H}_{FG}$ or ${\cal L}_{FG}$. For that
purpose, for each element in ${\cal H}_{FG}$ and each element in
${\cal L}_{FG}$ we introduce a map to graphs:
$$ \overline{Z_{X}}= X,\overline{ \delta_{X}}=
X.$$

\noindent Further, we write $\Delta(X)=\sum_i
X^\prime_{(i)}\otimes X^{\prime\prime}_{(i)}$ for the coproduct in
the Hopf algebra ${\cal H}_{FG}$.

\noindent The Lie algebra ${\cal L}_{FG}$ gives rise to two
representations acting as derivations on the Hopf algebra ${\cal
H}_{FG}$:

$$ Z^+_\Gamma\times \delta_X=\delta_{X\star\Gamma}$$
and
$$ Z^-_\Gamma\times\delta_X=\sum_i \langle Z^+_\Gamma,X^\prime_{(i)}\rangle X^{\prime\prime}_{(i)}.$$
Furthermore, any term in the coproduct of a 1PI graph $\Gamma$
determines gluing data $G_i$ such that
$$\Gamma=\Gamma^{\prime\prime}_{(i)}\star_{G_i}\Gamma^\prime_{(i)},
\forall i.$$ Here, $G_i$ specifies vertices in
$\Gamma^{\prime\prime}_{(i)}$ and bijections of their types with
the elements of $\Gamma^{\prime}_{(i)}$ such that $\Gamma$ is
regained from its parts:
$$\pica .$$
The first line gives a term $(i)$ in the coproduct, decomposing
this graph into its only divergent subgraph (assuming we have
chosen $\phi^3$ in six dimensions, say) and the corresponding
cograph, the second line shows the gluing  $G_i$ for this term, in
this example .

\noindent We want to understand the commutator
$$[Z_{\Gamma_1}^+,Z_{\Gamma_2}^-] ,$$
acting as a derivation on the Hopf algebra element $\delta_X$. To
this end introduce
$$Z_{[\Gamma_1,\Gamma_2]}\times\delta_X=\sum_i \langle Z_{\Gamma_2}^+,X^\prime_{(i)}\rangle
X^{\prime\prime}_{(i)}\star_{G_i}\Gamma_1.$$ Here, the gluing
operation $G_i$ still acts such that each topologically different
graph is generated with unit multiplicity.
$$\picb .$$
Note that if ${\bf res}(\Gamma_1)\not\sim {\bf res}(\Gamma_2)$,
$Z_{[\Gamma_1,\Gamma_2]}\times\delta_X$ vanishes, as the existence
of a bijection between edges adjacent to $\Gamma_2$ in $X$ and
$\Gamma_{1,{\rm ext}}^{[1]}$ demands the compatibility of the
residues of $\Gamma_1,\Gamma_2$.

\noindent Let $X,Y$ be related as
$\overline{Z_{[\Gamma_1,\Gamma_2]}\times \delta_X}=Y$, for a 1PI
graph $X$. Then, $Y$ is a sum of say $k$ 1PI graphs. We
immediately have thanks to our gluing conventions\\[2mm] {\bf
Proposition.} {\it
 $\overline{Z_{[\Gamma_2,\Gamma_1]}\times
\delta_Y}=k\;X$.\\[2mm]} Let us now consider
$$ [Z_{[\Gamma_1,\Gamma_2]},Z_{[\Gamma_3,\Gamma_4]}]\times\delta_X.$$

\noindent We first define
$$ Y_{234}:=\{Y\in {\cal H}_{FG,L}|\langle Z_{\Gamma_2},Z_{[\Gamma_3,\Gamma_4]}\times\delta_Y\rangle=1\}$$
and
$$ Y_{412}:=\{Y\in {\cal H}_{FG,L}|\langle Z_{\Gamma_4},Z_{[\Gamma_1,\Gamma_2]}\times\delta_Y\rangle=1\}.$$
Let $\Delta_\Gamma:{\cal H}_{FG}\to {\cal H}_{FG}\otimes {\cal
H}_{FG}$ be the map
\begin{equation}
X\to \sum_i X^\prime_{(i)}\otimes
[X^{\prime\prime}_{(i)}\star_{G_i}\Gamma]
\end{equation}
and let us write $\partial_2$ for the map $X\to \langle
Z^+_{\Gamma_2},X\rangle$.  Then,
$$Z_{[\Gamma_1,\Gamma_2]}\times\delta_X=
(\partial_{2}\otimes {\rm id})\circ\Delta_{\Gamma_1}$$ which
justifies the shorthand notation $ 1^+\partial_2 X$ for the above.

\noindent Then, the desired commutator is $$[1^+\partial_2
3^+\partial_4- 3^+\partial_4 1^+\partial_2]X.$$ Let us consider
$1^+\partial_2 3^+\partial_4 X$ first. We want to compare it with
$1^+3^+\partial_{2,4}X$. These are the terms generated by
shrinking $\Gamma_2,\Gamma_4$ at disjoint places, and gluing
$\Gamma_1$ for the residue of $\Gamma_2$, and $\Gamma_3$ for the
residue of $\Gamma_4$.

\noindent What we now need to know is the
commutator $1^+[\partial_2,3^+]\partial_4$. There are two cases:\\
i) $\Gamma_2$ is a proper subgraph of $\Gamma_3$,
$\Gamma_2\subset\Gamma_3$. Then,
$$1^+\partial_2 3^+\partial_4 X=(1^+\partial_2 \Gamma_3)^+\partial_4 X+1^+3^+\partial_{2,4}X.$$
ii) $\Gamma_2\not\subset\Gamma_3$. Then, for any $X_{(i)}^\prime$
such that $X_{(i)}^\prime=Y$, $Y\in Y_{234}$, we have a
contribution as $3^+\partial_4 Y=\Gamma_2$, and by the previous
proposition, $\Gamma_2= 4^+\partial_3 Y$. Hence
$$ 1^+\partial_2 3^+\partial_4 X=1^+\partial_{4^+\partial_3 2}X+1^+3^+\partial_{2,4}X.$$

\noindent Consider now $ 3^+\partial_4 1^+\partial_2 X$.
Similarly, we find two cases:\\ i) $\Gamma_4$ is a proper subgraph
of $\Gamma_1$, $\Gamma_4\subset\Gamma_1$. Then,
$$3^+\partial_4 1^+\partial_2 X=(3^+\partial_4 \Gamma_1)^+\partial_2 X+3^+1^+\partial_{4,2}X.$$
ii) $\Gamma_4\not\subset\Gamma_1$. Then, for any $X_{(i)}^\prime$
such that $X_{(i)}^\prime=Y$, $Y\in Y_{412}$, we have a
contribution as $1^+\partial_2 Y=\Gamma_4$, and by the proposition
again, $\Gamma_4= 2^+\partial_1 Y$. Hence
$$ 3^+\partial_4 1^+\partial_2 X=3^+\partial_{2^+\partial_1 4}X+3^+1^+\partial_{4,2}X.$$

\noindent As $$1^+3^+\partial_{2,4}X= 3^+1^+\partial_{4,2}X,$$ we
get for the commutator, returning to the full fledged notation,
\begin{eqnarray*}
[Z_{[\Gamma_1,\Gamma_2]},Z_{[\Gamma_3,\Gamma_4]}] & = &
+Z_{[\overline{Z_{[\Gamma_1,\Gamma_2]}\times\delta_{\Gamma_3}},\Gamma_4]}
-Z_{[\Gamma_3,\overline{Z_{[\Gamma_2,\Gamma_1]}\times\delta_{\Gamma_4}}]}\\
& & -
Z_{[\overline{Z_{[\Gamma_3,\Gamma_4]}\times\delta_{\Gamma_1}},\Gamma_2]}
+Z_{[\Gamma_1,\overline{Z_{[\Gamma_4,\Gamma_3]}\times\delta_{\Gamma_2}}]}\\
 & &
 -\delta^K_{\Gamma_2,\Gamma_3}Z_{[\Gamma_1,\Gamma_4]}+\delta^K_{\Gamma_1,\Gamma_4}Z_{[\Gamma_2,\Gamma_3]}.
\end{eqnarray*}
Let us check that this bracket fulfills a Jacobi identity.
Equivalently, we can check that $$Z_{[\Gamma_1,\Gamma_2]}\star
Z_{[\Gamma_3,\Gamma_4]}:=Z_{[\overline{Z_{[\Gamma_1,\Gamma_2]}\times\delta_{\Gamma_3}},4]}+
Z_{[\Gamma_1,\overline{Z_{[\Gamma_4,\Gamma_3]}\times\delta_{\Gamma_2}}]}$$
defines a right or left pre-Lie product. Indeed, we find,
returning to our shorthand notation:
\begin{eqnarray*}
 & & (1^+\partial_2 3^+\partial_4)5^+\partial_6-1^+\partial_2(3^+\partial_45^+\partial_6)\\
 & = & +(1^+(\partial_2 3))^+\partial_45^+\partial_6+
 1^+\partial_{4^+\partial_3 2}5^+\partial_6\\
 & & -1^+\partial_2 (3^+\partial_4 5)^+\partial_6-1^+\partial_2
 3^+\partial_{6^+\partial_54}\\
  & = & +\underbrace{((1^+\partial_2 3)^+\partial_4
  5)^+\partial_6}_{b_1}
  +
  \underbrace{(1^+\partial_2 3)^+\partial_{6^+\partial_5 4}}_a\\
& &
+\underbrace{(1^+\partial_{4^+\partial_32}5)^+\partial_6}_{b_2}+\underbrace{1^+\partial_{6^+\partial_5
(4^+\partial_3 2)}}_{c_3}\\
 & & - \underbrace{(1^+\partial_2 (3^+\partial_4
 5))^+\partial_6}_{b_3}
  -
  \underbrace{1^+\partial_{6^+\partial_{3^+\partial_4 5} 2}}_{c_1}  \\
 & & -
 \underbrace{(1^+\partial_2 3)^+\partial_{6^+\partial_5 4}}_a
 -
 \underbrace{1^+\partial_{(6^+\partial_5 4)^+\partial_3 2}}_{c_2}
\end{eqnarray*}
The two ''a'' terms cancel, while the terms $b_1,b_2,b_3$ add up
to a contribution $(1^+3^+\partial_{2,4} 5)^+ \partial_6$ which is
symmetric under exchange of the index pair $(1,2)$ with $(3,4)$.
This term only contributes when $\Gamma_2$ appears as a subgraph
of $\Gamma_5$. The terms $c_1,c_2,c_3$ add up to a contribution
$1^+\partial_{6^+4^+\partial_{5,3} 2}$ which only contributes when
$\Gamma_5$ appears as a subgraph of $\Gamma_2$, and is symmetric
under exchange of the index pair $(3,4)$ with $(5,6)$. The
$b_i$-terms and the $c_i$ terms are mutually exclusive.
Furthermore, when the $b_i$ terms contribute, we get a right
pre-Lie product, while when the $c_i$ terms contribute, we get a
left pre-Lie product. In all cases, we then fulfill the
Jacobi identity.~$\|$\\[5mm]
Hence, we have established the following theorem:\\[2mm] {\bf Theorem.}
{\it For all 1PI graphs $\Gamma_i$, s.t.~${\bf res}(\Gamma_1)={\bf
res}(\Gamma_2)$ and ${\bf res}(\Gamma_3)={\bf res}(\Gamma_4)$, the
bracket \begin{eqnarray*}
[Z_{[\Gamma_1,\Gamma_2]},Z_{[\Gamma_3,\Gamma_4]}] & = &
+Z_{[\overline{Z_{[\Gamma_1,\Gamma_2]}\times\delta_{\Gamma_3}},\Gamma_4]}
-Z_{[\Gamma_3,\overline{Z_{[\Gamma_2,\Gamma_1]}\times\delta_{\Gamma_4}}]}\\
& & -
Z_{[\overline{Z_{[\Gamma_3,\Gamma_4]}\times\delta_{\Gamma_1}},\Gamma_2]}
+Z_{[\Gamma_1,\overline{Z_{[\Gamma_4,\Gamma_3]}\times\delta_{\Gamma_2}}]}\\
 & &
 -\delta^K_{\Gamma_2,\Gamma_3}Z_{[\Gamma_1,\Gamma_4]}+\delta^K_{\Gamma_1,\Gamma_4}Z_{[\Gamma_3,\Gamma_2]}.
\end{eqnarray*}
defines a Lie algebra of derivations acting on the Hopf algebra
${\cal H}_{FG}$ via $$ Z_{[\Gamma_i,\Gamma_j]}\times
\delta_X=\sum_I\langle Z^+_{\Gamma_2},\delta_{X^\prime_{(i)}}
\rangle \delta_{X^{\prime\prime}_{(i)}\star_{G_i}\Gamma_1},$$
where the gluing data $G_i$ are normalized as before.\\[2mm]} The
Kronecker $\delta^K$ terms just eliminate the overcounting when
combining all cases in a single equation.

\noindent  We note that $Z_{[\Gamma,\Gamma]}\times
\delta_X=k_\Gamma \delta_X$, where $k_\Gamma$ is the number of
appearances of $\Gamma$ in $X$ and where we say that a graph
$\Gamma$ appears $k$ times in $X$ if $k$ is the largest integer
such that
$$ \langle \Gamma^k\otimes {\rm id},\Delta(\delta_X)\rangle$$ is non-vanishing.

\noindent  Furthermore, we note that $I:
Z_{[\Gamma_1,\Gamma_2]}\to Z_{[\Gamma_2,\Gamma_1]}$ is an
anti-involution such that
$$I([Z_{[\Gamma_1,\Gamma_2]},Z_{[\Gamma_3,\Gamma_4]}])=
-[I(Z_{[\Gamma_1,\Gamma_2]}),I(Z_{[\Gamma_3,\Gamma_4]})],$$ by
inspection. We have
$$[Z_{[\Gamma_1,\Gamma_2]},Z_{[\Gamma_2,\Gamma_1]}]=Z_{[\Gamma_1,\Gamma_1]}-Z_{[\Gamma_2,\Gamma_2]}.$$
Further structural analysis is left to future work.

\noindent By construction, we have\\[2mm] {\bf Proposition.} {\it
$$Z^+_\Gamma\equiv Z_{[\Gamma,{\bf res}(\Gamma)]},$$ $$Z^-_\Gamma\equiv
Z_{[{\bf res}(\Gamma),\Gamma]}.$$ } \noindent Also, we immediately
conclude\\[2mm]
{\bf Corollary.} {\it
$[Z^-_X,Z^-_Y]=-Z^-_{\overline{[Z^+_X,Z^+_Y]}}.$ }

\noindent Finally, we get the desired commutator\\[2mm] {\bf
Corollary.} {\it   {\small
\begin{eqnarray*}
[Z_{[\Gamma_1,{\bf res}(\Gamma_1)]},Z_{[{\bf
res}(\Gamma_2),\Gamma_2]}] & = & +Z_{[\overline{Z_{[\Gamma_1,{\bf
res}(\Gamma_1)]}\times\delta_{{\bf res}(\Gamma_2)}},\Gamma_2]}
-Z_{[{\bf res}(\Gamma_2),\overline{Z_{[{\bf res}(\Gamma_1),\Gamma_1]}\times\delta_{\Gamma_2}}]}\\
& & - Z_{[\overline{Z_{[{\bf
res}(\Gamma_2),\Gamma2]}\times\delta_{\Gamma_1}},{\bf
res}(\Gamma_1)]}
+Z_{[\Gamma_1,\overline{Z_{[\Gamma_2,{\bf res}(\Gamma_2)]}\times\delta_{{\bf res}(\Gamma_1)}}]}\\
 & &
 -\delta^K_{{\bf res}(\Gamma_1),{\bf res}(\Gamma_2)}Z_{[\Gamma_1,\Gamma_2]}+\delta^K_{\Gamma_1,\Gamma_2}Z_{[{\bf res}(\Gamma_2),{\bf res}(\Gamma_1)]}\\
  & = & \delta^K_{{\bf res}(\Gamma_1),{\bf res}(\Gamma_2)}Z_{[\Gamma_1,\Gamma_2]}+\delta^K_{\Gamma_1,\Gamma_2}Z_{[{\bf res}(\Gamma_2),{\bf res}(\Gamma_1)]}\\
   & & - Z_{[{\bf res}(\Gamma_2),\overline{Z_{[{\bf res}(\Gamma_1),\Gamma_1]}\times\delta_{\Gamma_2}}]}
   - Z_{[\overline{Z_{[{\bf
res}(\Gamma_2),\Gamma2]}\times\delta_{\Gamma_1}},{\bf
res}(\Gamma_1)]}\\
  & = & \delta^K_{{\bf res}(\Gamma_1),{\bf res}(\Gamma_2)}Z_{[\Gamma_1,\Gamma_2]}+\delta^K_{\Gamma_1,\Gamma_2}Z_{[{\bf res}(\Gamma_2),{\bf res}(\Gamma_1)]}\\
   & & - Z^-_{\overline{Z_{[{\bf res}(\Gamma_1),\Gamma_1]}\times\delta_{\Gamma_2}}}
   - Z^+_{\overline{Z_{[{\bf
res}(\Gamma_2),\Gamma2]}\times\delta_{\Gamma_1}}}.
\end{eqnarray*}}
} \noindent We can now make contact with derivations in the Hopf
algebra of rooted trees. Let us consider the Hopf algebra of
iterated one-loop self-energies in massless Yukawa theory in four
dimensions. There is a one-to-one correspondence $\Theta$ between
iterated one-loop  fermion self-energy graphs and undecorated
rooted trees:
$$\picgt .$$
Let $\Gamma_2,\Gamma_3$ be arbitrary such fermion self-energy
graphs and let $\Gamma_4$ be the one-loop self-energy graph, and
$\Gamma_1$ be its residue, a two-point vertex with two fermionic
external legs.

\noindent Note that
$${\bf res}(\Gamma_4)={\bf res}(\Gamma_2)={\bf
res}(\Gamma_3)=\Gamma_1.$$

\noindent The isomorphism $\Theta$ to undecorated rooted trees
delivers the previous result on undecorated rooted trees: Indeed,
$$\Theta(Z_{[\Gamma_3,\Gamma_4]})=N(\Theta(\Gamma_3)),$$ and
$$\Theta(Z_{[\Gamma_1,\Gamma_2]})=M(\Theta(\Gamma_2)).$$
We have, using the previous theorem,
\begin{eqnarray*}
\Theta([Z_{[\Gamma_1,\Gamma_2]},Z_{[\Gamma_3,\Gamma_4]}]) & =&
[M(\Theta(\Gamma_2)),N(\Theta(\Gamma_3))]\\
 & = & \Theta\left[
 Z_{[
 \overline{Z_{
 [{\bf
 res}(\olp),\Gamma_2]\times\delta_{\Gamma_3}}},\olp]}\right.\\
 & &
 \left. +Z_{[{\bf
 res}(\olp),
 \overline{Z_{
 [\olp,\Gamma_3]\times\delta_{\Gamma_2}}}]}\right. \\
  & &
\left.-\delta^K_{\Gamma_2,\Gamma_3}Z_{[{\bf
res}(\olp),\olp]}\right],
\end{eqnarray*}
in accordance with the results of the previous section. We used
the fact that the residue of a graph contains no subgraph,
$$ Z_{[\Gamma_3,\olp]}\times \delta_{{\bf res}(\olp)}=0,$$
and that
$$Z_{[\Gamma_2,{\bf res}(\olp)]}\times\delta_{\olp}=0.$$

\noindent The above uses naturally growth by identifying the root
of a tree with any feet of another tree. We can also work out from
our general results the commutator of other derivations, using,
for example, natural growth by connecting with an extra edge the
root of a tree to all the vertices of another one.
\section*{Conclusions}
We only considered the Lie algebra aspect for Feynman graphs. A
bicrossed structure can be constructed as well, say by enlarging
${\cal H}_{FG}$ to ${\cal H}_{FGG}$ using appropriate insertion of
subgraphs as a natural growth.

\noindent The algebraic structures here provided cover all
operations which one encounters in the perturbative expansion of a
quantum field theory: insertion and elimination of subgraphs.
While the construction of local counterterms demands the
elimination of subgraphs $\gamma$ by ${\bf res}(\gamma)$ on the
expense of multiplication with their counterterms $S_R(\gamma)$
\cite{RHI}, the Dyson--Schwinger quantum equations of motions
demand that any local interaction, described by a vertex $v$, can
as well be mediated by any graph $\Gamma$ with ${\bf
res}(\Gamma)=v$, and hence the insertion of $\Gamma$ for $v$ in
all possible ways determines naturally the series of Feynman
graphs providing a fixpoint for those equations.

\end{document}